\documentclass[aps,prl,preprint,groupedaddress]{revtex4}

\usepackage{graphicx}
\usepackage{dcolumn}
\usepackage{bm}
\usepackage{amsmath}
\usepackage{color}
\begin{document}

\title{Stabilizing effects of resistivity on low-$n$ edge localized modes in NSTX}

\author{Debabrata Banerjee}
%
\author{Ping Zhu}
\email[]{pzhu@ustc.edu.cn}
\affiliation{CAS Key Laboratory of Geospace Environment and Department of Modern Physics, University of Science and Technology of China, Hefei, Anhui 230026, China}
\affiliation{Department of Engineering Physics, University of Wisconsin-Madison, Madison, Wisconsin 53706, USA}

\author{Rajesh Maingi}
\affiliation{Princeton Plasma Physics Laboratory, P.O. Box $451$, Princeton, New Jersey 08543, USA}
\date{\today}
             
\begin{abstract}
The stabilizing effects of enhanced edge resistivity on the low-$n$ edge localized modes (ELMs) are reported for the first time in the context of ELM suppression in H-mode discharge due to lithium-conditioning in the National Spherical Torus Experiment (NSTX). Here $n$ is the toroidal mode number. Linear stability analysis of the corresponding experimental equilibrium suggests that the change in the equilibrium plasma density profile alone due to lithium-conditioning may be insufficient for a complete suppression of ELMs. The enhanced resistivity due to the increased effective electric charge number $Z_{\rm eff}$ after lithium-conditioning can account for additional stabilization effects that contribute to robust ELM suppression. Remarkably, such a stabilizing effect of enhanced edge resistivity on the low-$n$ ELMs only exists when two-fluid effects are considered in the MHD model.
\end{abstract}

\pacs{}

\maketitle

 The edge localized mode (ELM) has been a primary concern to the next generation of long-pulse, steady-state burning plasma experiments such as ITER~\cite{loarte2007,langnf} and CFETR~\cite{vschan}. This is because ELMs play a central role in the periodic loss of particles and power from the collapse of edge pedestal in the high confinement regime, i.e. H-mode of tokamaks~\cite{wagner82,zohm96,connor98}. Apart from the deterioration in plasma confinement, they cause physical damage to divertor plates and other plasma facing components through repetitive heat loads and particle deposition~\cite{federici}. One of the effective ELM suppression and mitigation method has been the injection of low-$z$ impurity particles (e.g, lithium) into tokamak during or immediately prior to the discharge~\cite{huprl,maingiprl,osbornenf}. This method was first successfully applied in NSTX through lithium coated divertor plates and ELM-free H-mode discharges have been achieved~\cite{maingiprl,maingiprl2}. Then, combined techniques of real time lithium-injection and prior discharge lithium-coating has been effectively applied to the Experimental Advanced Superconducting Tokamak (EAST) to produce $20$ sec long ELM-free H-mode discharge~\cite{huprl}. More detailed understanding of this method is desirable in order to design its effective implementation towards controlling ELMs in ITER.

A systematic study of H-mode discharge with decreasing ELM frequency and amplitude has been carried out in NSTX after gradually increasing the thickness of lithium coatings on wall and divertor plates~\cite{mainginf,boyleppcf}. Lithium conditioning reduces plasma recycling which in turn helps to minimize core fueling from divertor recycling sources and thus the edge electron density in the scrape-off layer and the near-separatrix region are reduced. This process eventually moves inward the pedestal positions of electron density and temperature profiles away from separatrix and the profiles become widen in comparison to the pre-lithium discharges. Besides those profile modifications, toroidal rotation and effective atomic number ($Z_{\rm eff}$) substantially increase after lithium conditioning of plasma facing components, especially after ELM elimination. A series of recent reports from NSTX has documented in details of the effect and outcome of lithium-conditioning, including its potential relation to ELM-free H-mode. This letter presents a new explanation on the additional impurity induced stabilization of NSTX ELM-free discharge based on our independent findings of stabilizing effect of resistivity on the low-$n$ edge localized modes~\cite{debu,itpa}.

The equilibrium of NSTX ELMy discharge is predicted to be close to the kink/peeling unstable boundary from the analysis using the ideal magnetohydrodynamic (MHD) code ELITE with inclusion of the stabilizing effects from the diamagnetic drift and the finite Larmor radius. Following the same analysis, the post-lithium ELM-free case is found to be within the stable regime of the peeling-ballooning instability diagram. It is argued that due to the inward shift of the edge peak pressure gradient to a lower magnetic shear region and substantial decrease of pedestal pressure gradient, the ballooning components have become stable. The reduced profile peaking of the edge bootstrap current density has enabled the peeling components to become stable. Those arguments led to earlier conclusion that 
the whole stabilization scenario is mainly the consequence of equilibrium profile modification due to lithium-conditioning. 
%

\begin{figure}[htbp]
\includegraphics[width=7.5cm]{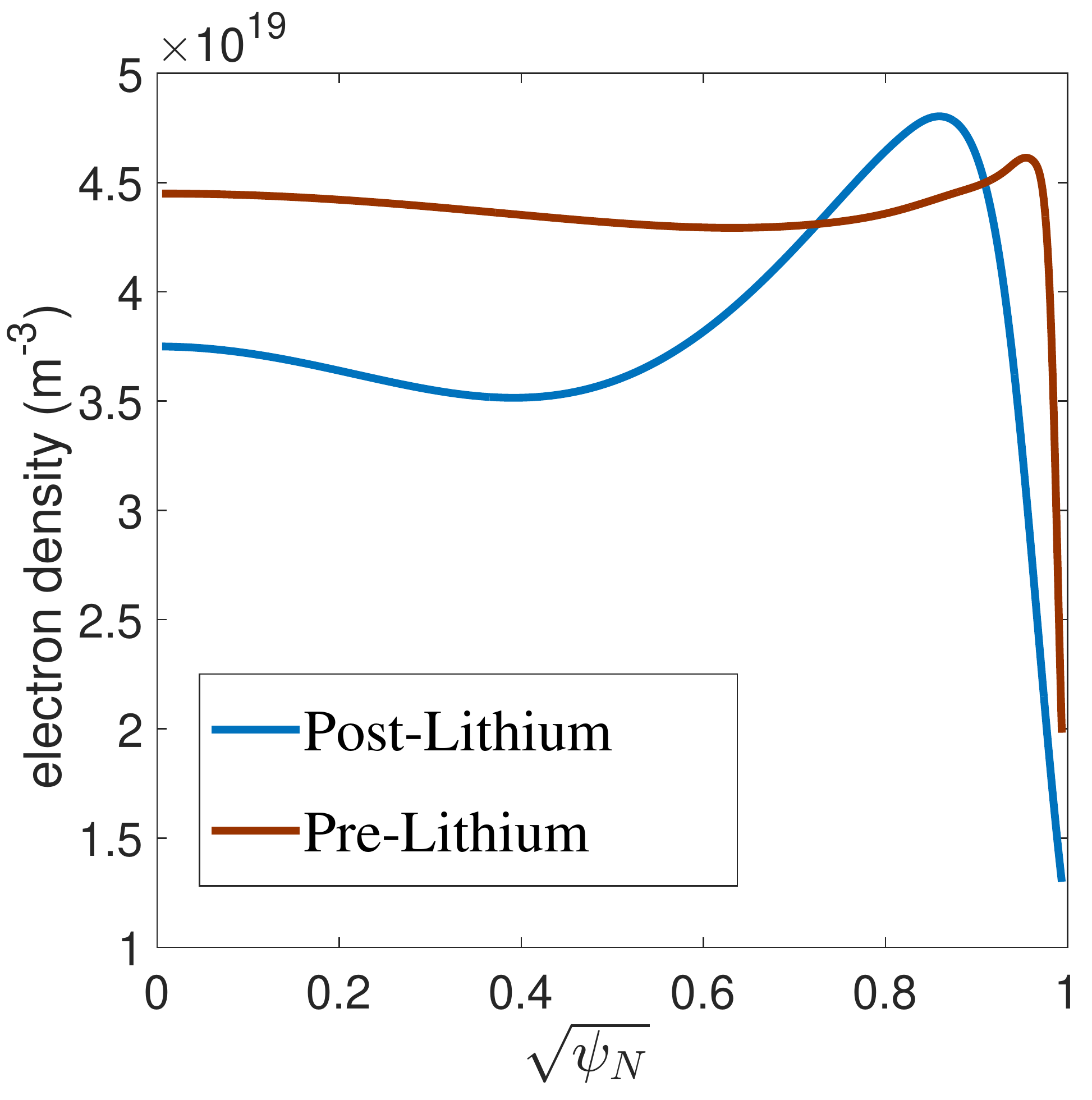}
~~\includegraphics[width=7.5cm]{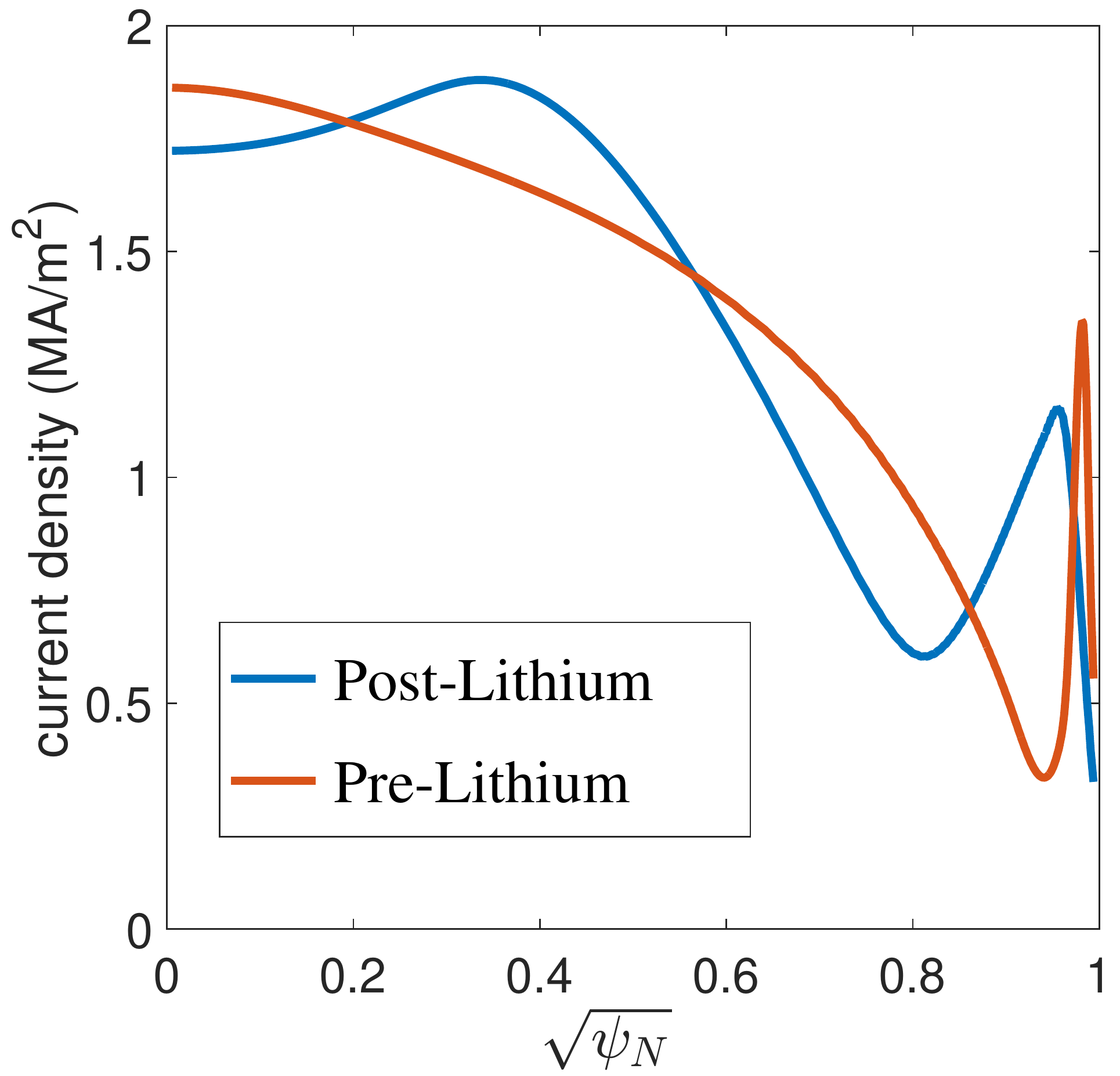}%
\caption{\label{equil} Electron density (left) and current density (right) radial profiles of NSTX ELM experiments. Blue solid lines correspond to post-lithium discharge and red dotted lines to the pre-lithium reference discharge. $\psi_{N}$ is the normalized poloidal flux function.}
 \end{figure}

This present letter contains the linear stability analysis results
from the first-principle based initial value extended MHD code NIMROD
\cite{sovinec04}, which has been benchmarked and verified for the
analysis of both ideal and non-ideal physical
processes~\cite{sovinec10,brennan,zhu13,burke,ebrahimi,izzo15,zhu08,jake}. The
equilibriums we study are from the pre-lithium reference ELMy H-mode
($\#129015$) discharge and the post-lithium ELM-free H-mode
($\#129038$) discharge from NSTX experiments~\cite{maingiprl}
(Fig.~\ref{equil}). The purpose of our analysis is to crosscheck the
cause behind the stabilization of ELMs, including the ideal MHD
effects from the equilibrium profile modifications and the non-ideal
MHD effects due to the increase of resistivity through $Z_{\rm eff}$
enhancement in the post-lithium discharge.  The change in resistivity
is not the only consequence of the $Z_{\rm eff}$ enhancement. In particular,
changing $Z_{\rm eff}$ can significantly change the bootstrap current
profile. Since the experimental equilibrium profiles are used in our
analysis, all profile changes, including those of plasma current from
changing $Z_{\rm eff}$, have been fully taken into account in our
calculations. Our findings suggest that the previous explanation on
the ELM suppression after lithium conditioning based purely on the profile
modification may be incomplete. The main new result of this letter is
that the enhancement of resistivity due to the increase of $Z_{\rm eff}$
after lithium usage, can play a direct and crucial role towards
ELM stabilization, in addition to the effects from pedestal profile
modifications.

The remainder of this letter is organized as follows. First, the main results of stabilization of the low-$n$ peeling-ballooning modes by enhanced resistivity with increase of $Z_{\rm eff}$ is described. Here $n$ is the toroidal mode number. Second, the convergence test with different crucial numerical parameters of simulation has been detailed. Third, similar stability analysis on the pre-lithium ELMy case has been shown, with clear indication of stabilizing effects of resistivity on the low-$n$ edge modes. Finally, conclusion is drawn along with discussions on the probable physical explanation.

The extended MHD equations used in our NIMROD calculations are: 
\begin{eqnarray}
\frac{\partial n}{\partial t} + \nabla \cdot \left( n {\bf u}\right) = 0 \\
m n \left( \frac{\partial}{\partial t} + {\bf u} \cdot \nabla \right) {\bf u} = {\bf J} \times {\bf B} - \nabla p - \nabla \cdot \overline{\Pi} \\
\frac{3}{2} \left( \frac{\partial}{\partial t} + {\bf u}_{\alpha} \cdot \nabla \right) T_{\alpha} = -n T_{\alpha} \nabla \cdot {\bf u}_{\alpha} - \nabla \cdot {\bf q}_{\alpha}\quad\quad (\alpha=i,e) \\
\frac{\partial {\bf B}}{\partial t} = - \nabla \times \left[ \eta {\bf J} - {\bf u}\times{\bf B} + \frac{1}{ne} \left({\bf J}\times{\bf B} - \nabla p_e \right)\right] \\
\mu_0 {\bf J} = \nabla \times {\bf B} ; ~~~~~~~~~~~~~~~ \nabla \cdot {\bf B} = 0
\end{eqnarray}
where {\bf u} is the center-of-mass flow velocity with particle density $n$ and ion mass $m$, $p$ is the combined pressure of electron ($p_e$) and ion ($p_i$), $\eta$ represents resistivity, ${\bf q}_{e,i}$
denote conductive heat flux vectors and $\overline{\Pi}$ is ion stress tensor including gyro-viscous components as described in the earlier reference~\cite{sovinec10}.

\begin{figure}[htbp]
\includegraphics[width=7.0cm]{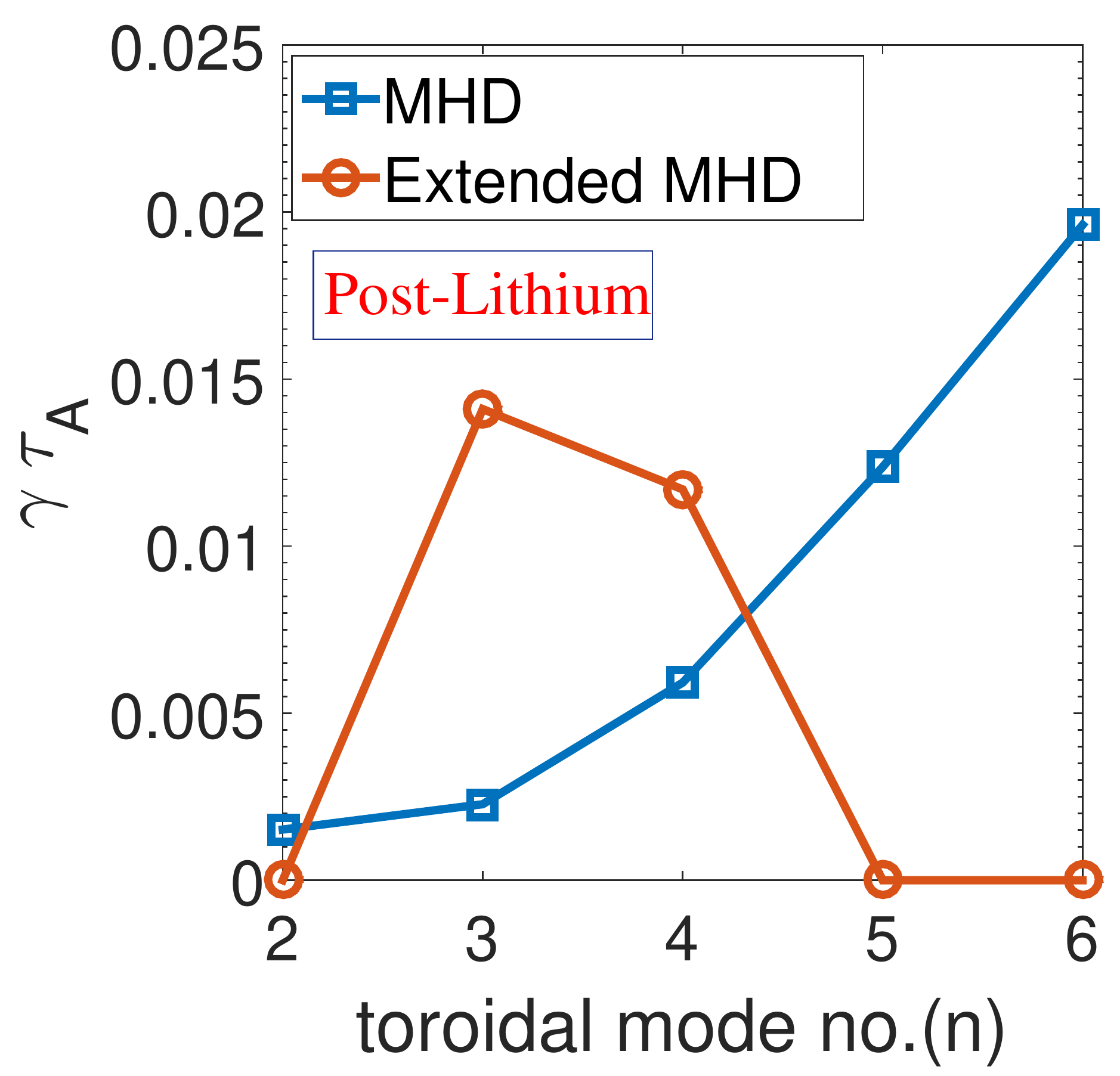}
~~\includegraphics[width=7.6cm]{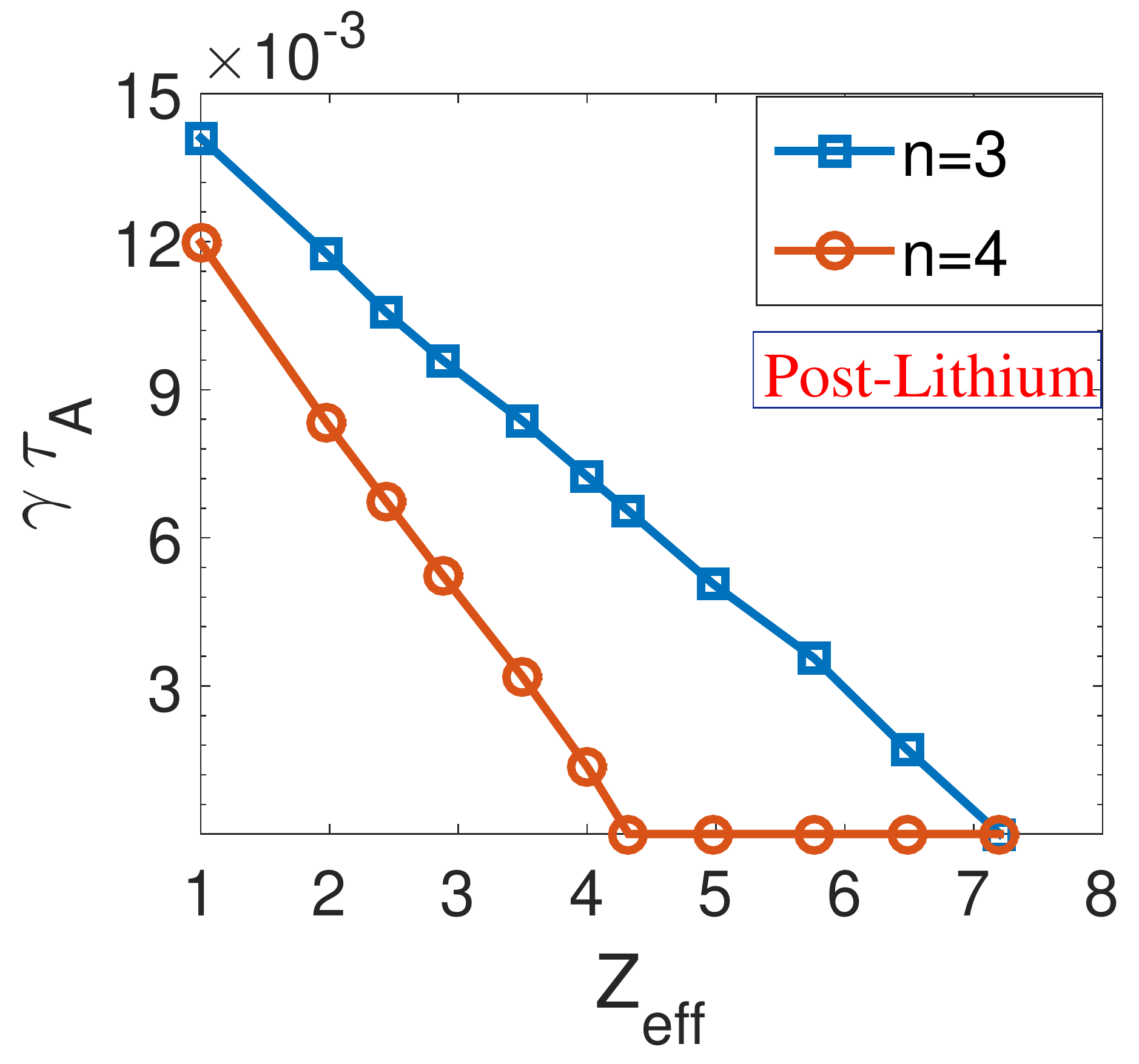}%
 \caption{\label{grpost} (Post-Lithium discharge $\#129038$)~Left: Linear growth rates as function of toroidal mode number from resistive MHD (blue circled curve) and extended MHD (red squared curve) calculations respectively using NIMROD for the post-lithium case. Right: Linear growth rates of $n=3,4$ modes as function of $Z_{\rm eff}$. Experimentally measured $Z_{\rm eff}$ at the edge ranges between 3.5 (at $\psi_N=0.9$) and 4.0 at ($\psi_N=0.7$) in the post-lithium discharge (see Fig.~$2$ of Ref.~\cite{mainginf}).}
\end{figure}

\begin{figure}[htbp]
\includegraphics[width=6.5cm]{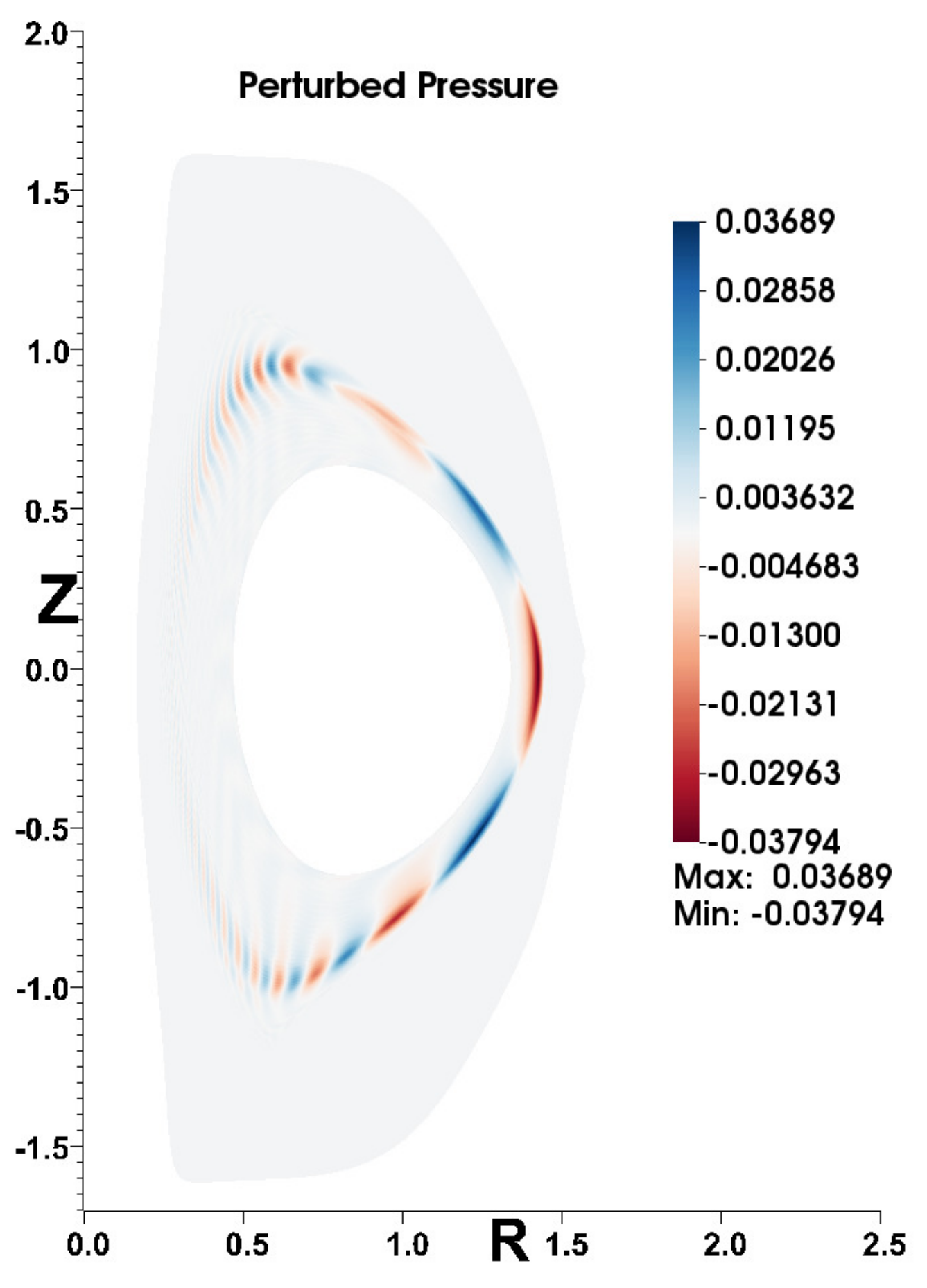}
~~~\includegraphics[width=6.5cm]{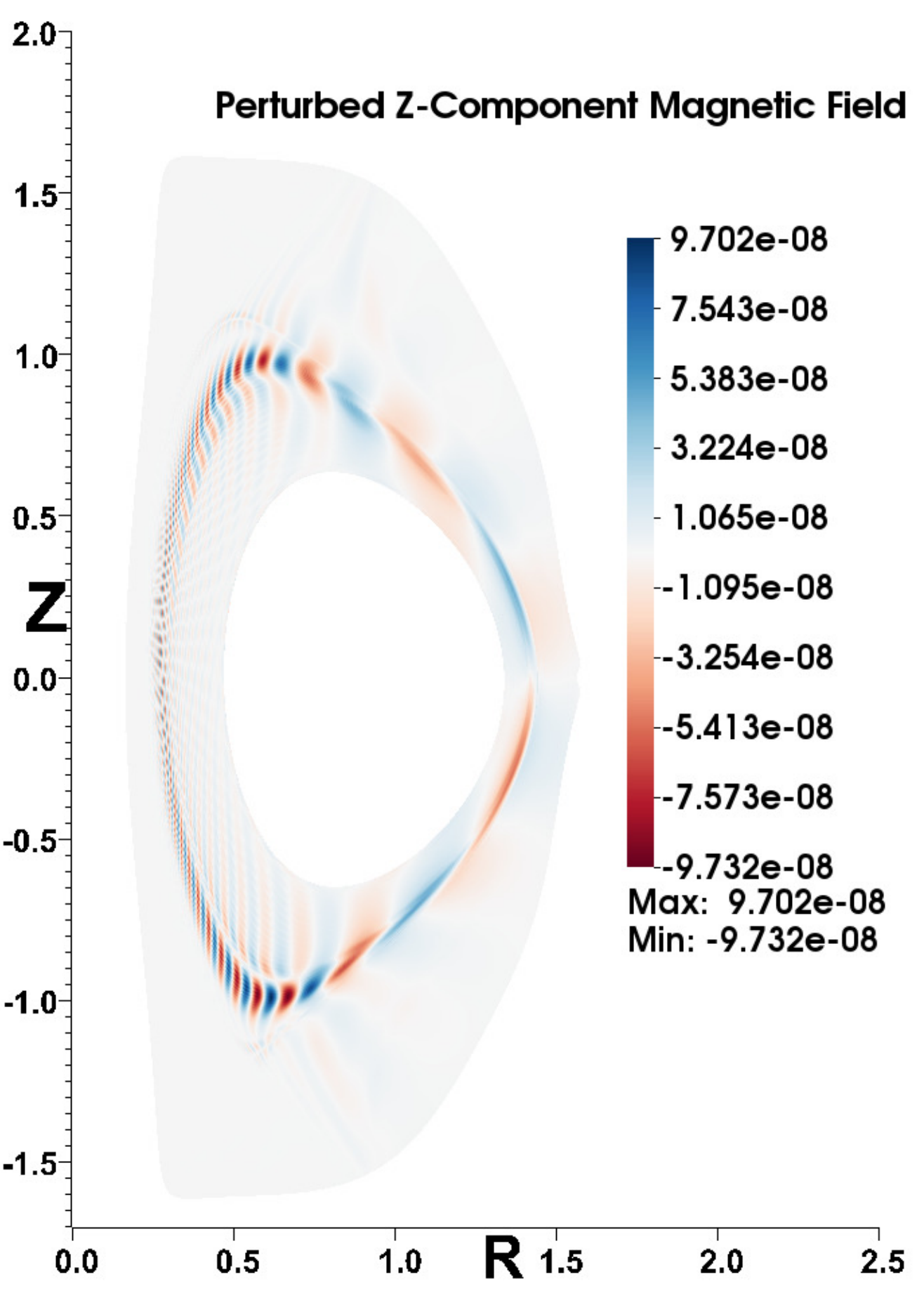} \\
~~~~~~~~~~~~~~~~~\\
~~~~~~~~~~~~~~~~~~~~\\
\includegraphics[width=6.5cm]{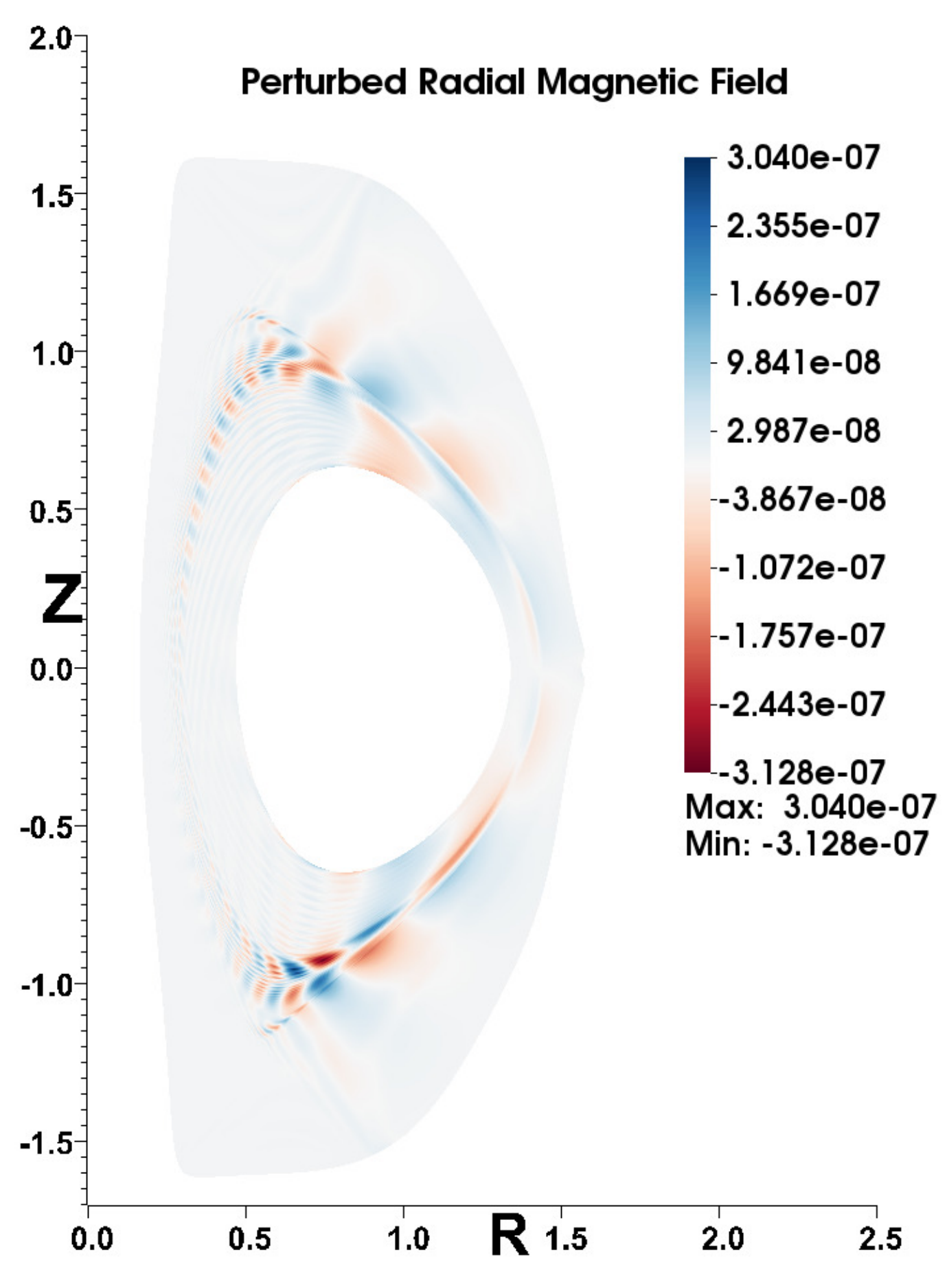}
~~~\includegraphics[width=6.5cm]{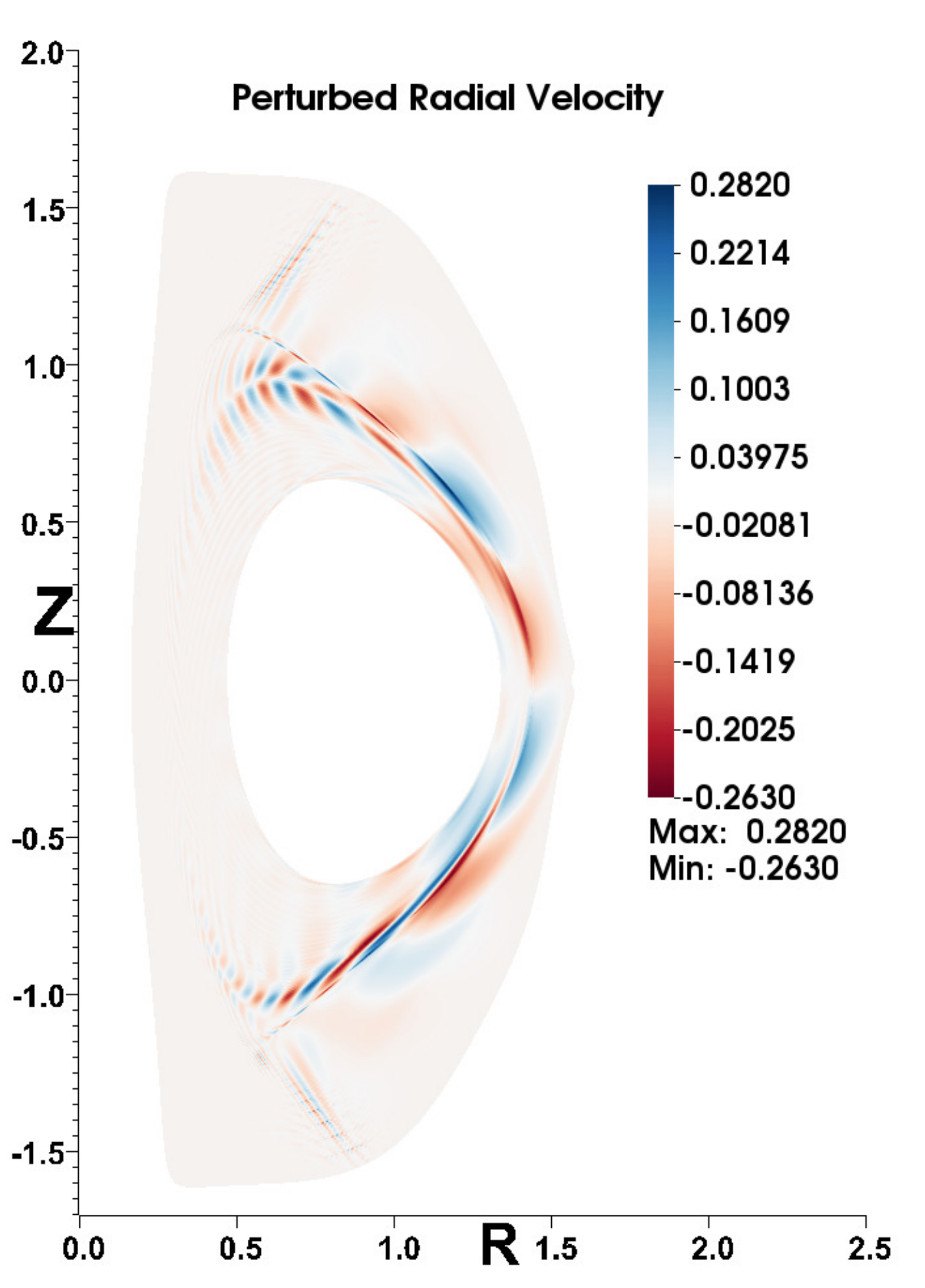}
 \caption{\label{contourpost} Contour plots of perturbed quantities pressure, $B_{z}$, $B_r$ and $u_r$ for $n=4$ mode for the post-lithium case. 
}
\end{figure}

\begin{figure}[htbp]
\includegraphics[width=7.5cm]{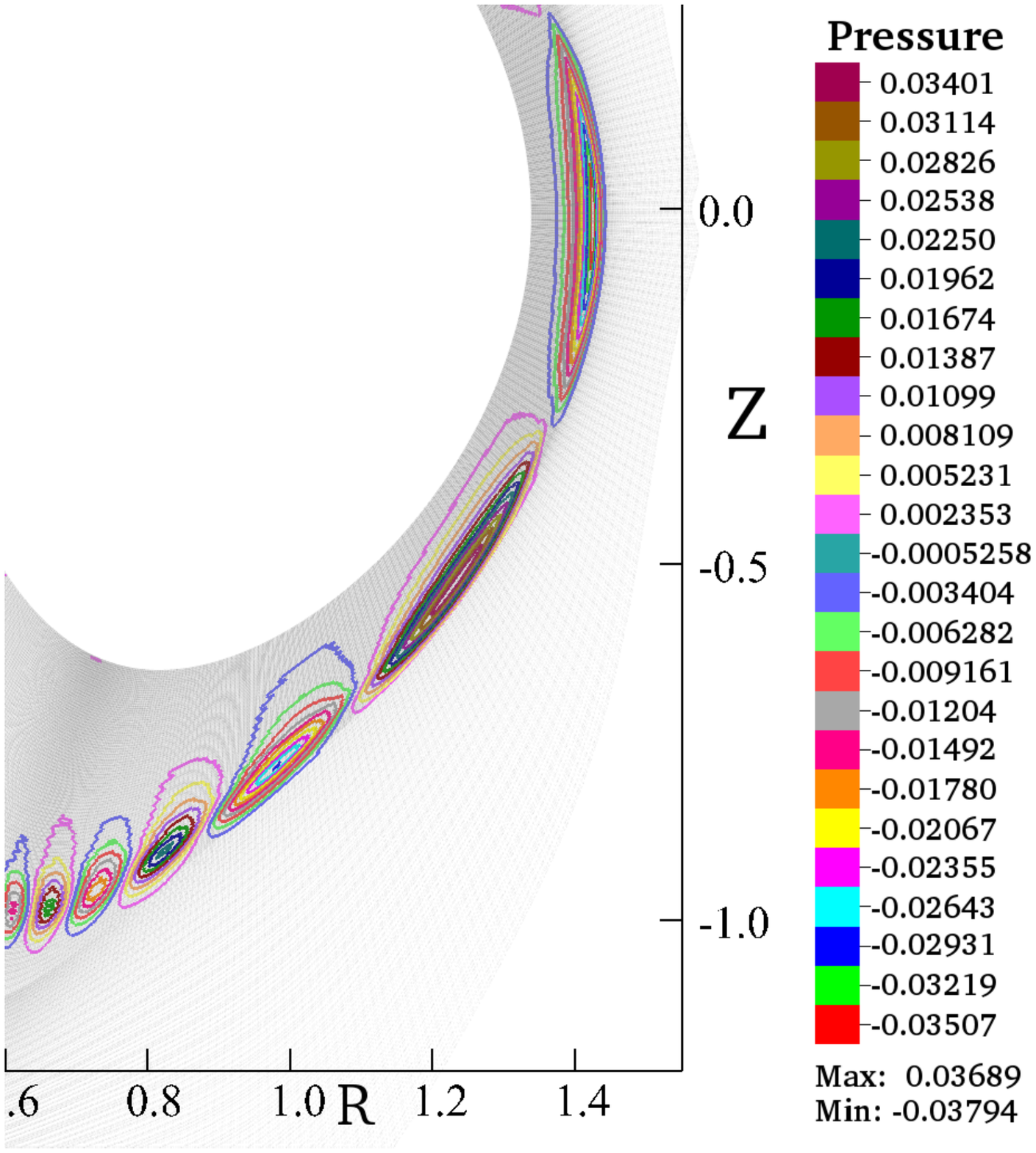}
~~~~\includegraphics[width=7.8cm]{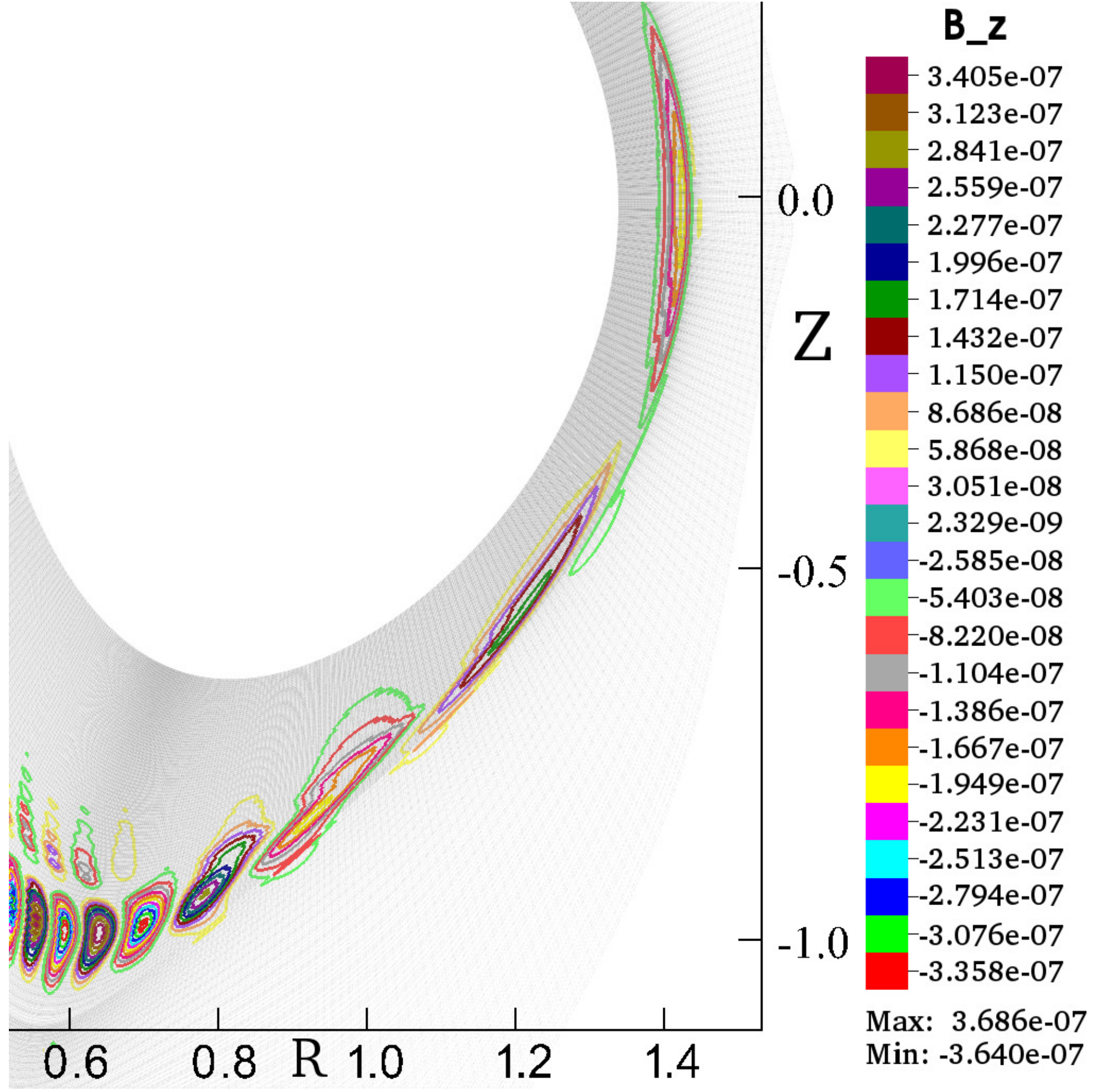}\\
~~~~~~~~~~\\
~~~~~~~~~~~~\\
~~~~~~~~~~~\\
\includegraphics[width=7.5cm]{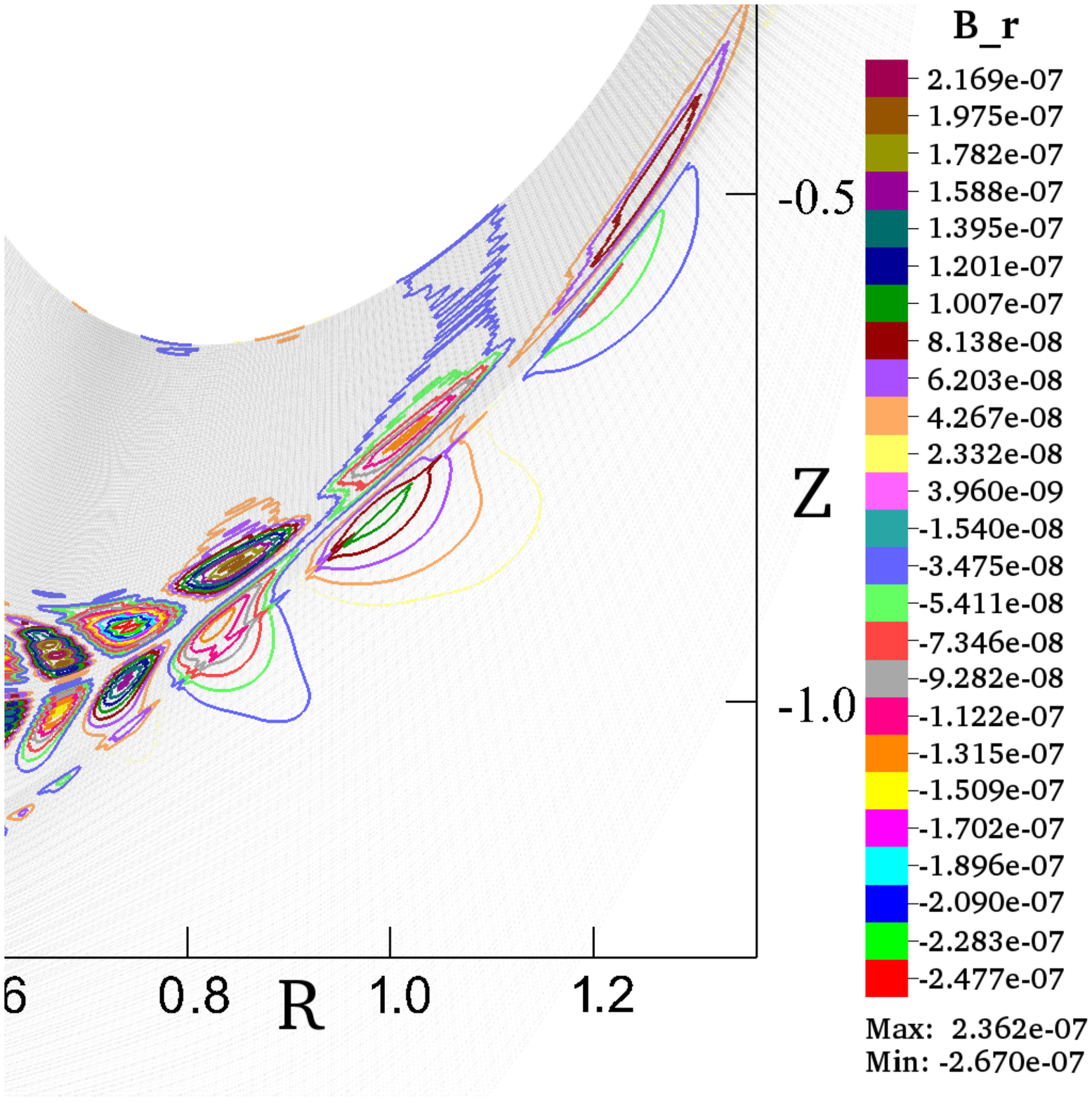}
~~~~\includegraphics[width=7.5cm]{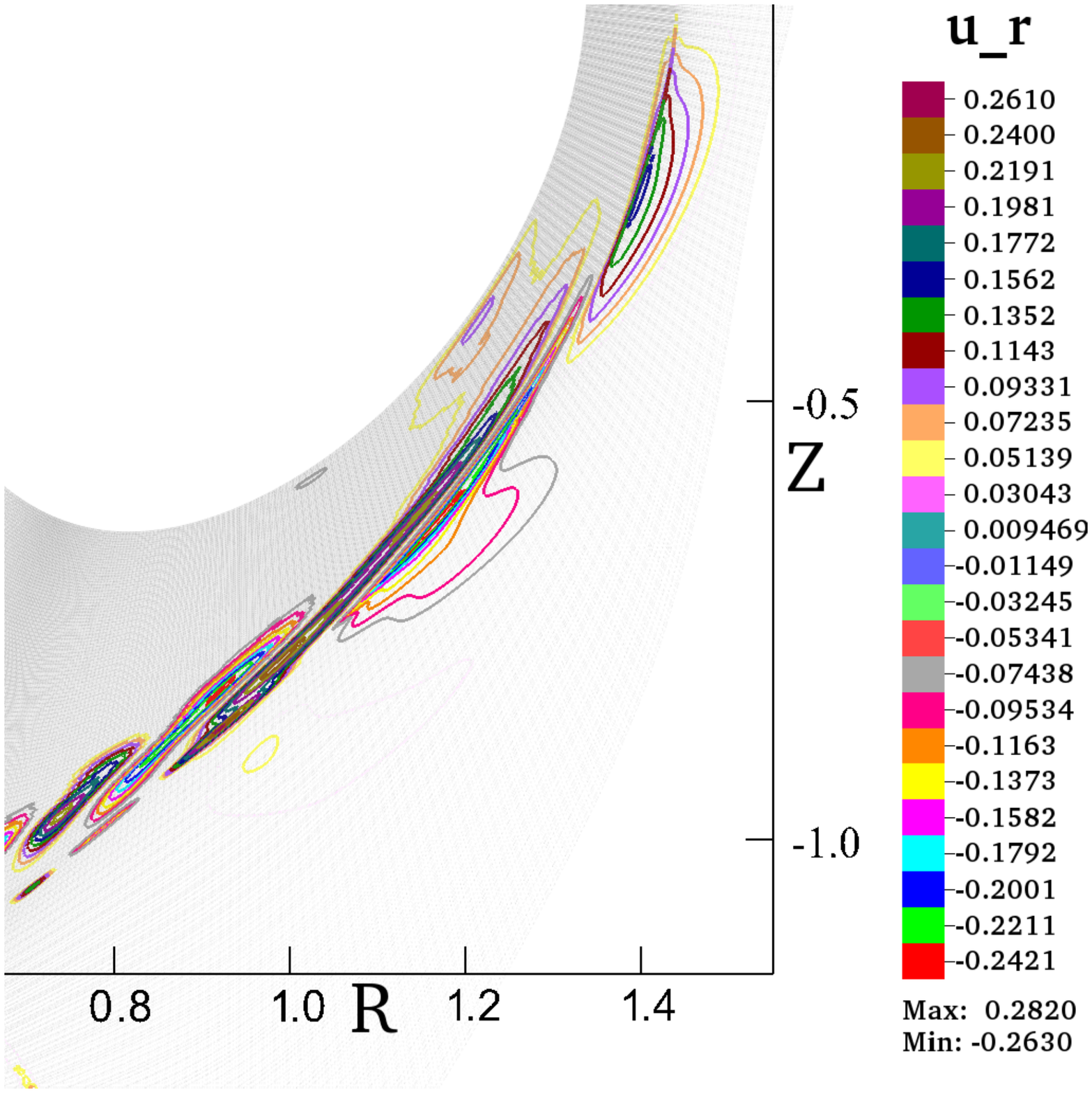}
\caption{\label{contourpost_zoom} Zoomed-in contour plots of perturbed quantities pressure, $B_{z}$, $B_r$ and $u_r$ for $n=4$ mode for the post-lithium case, showing detailed mode structure. 
}
\end{figure}

\begin{figure}[htbp]
\includegraphics[width=7cm]{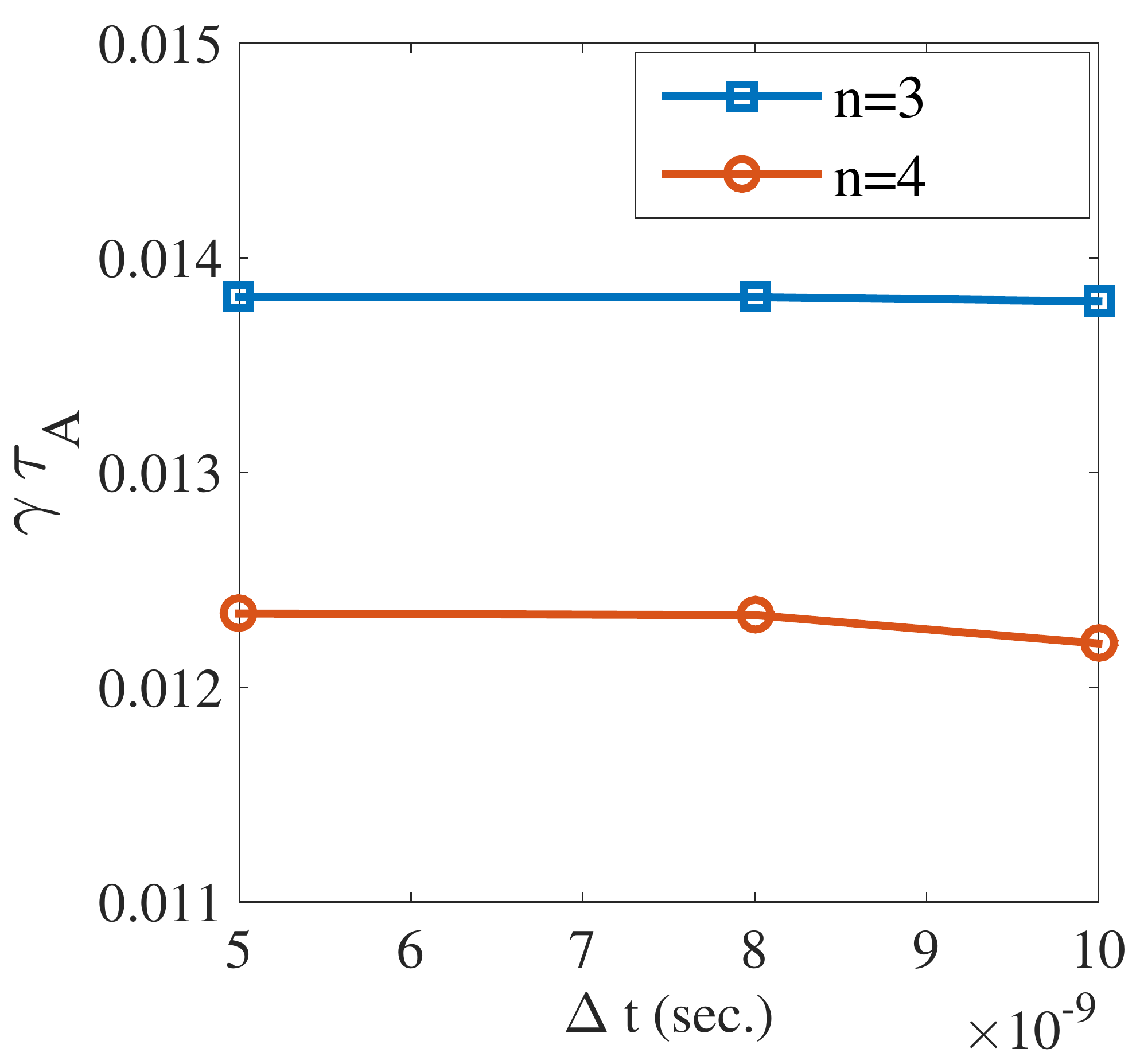}
~~\includegraphics[width=7cm]{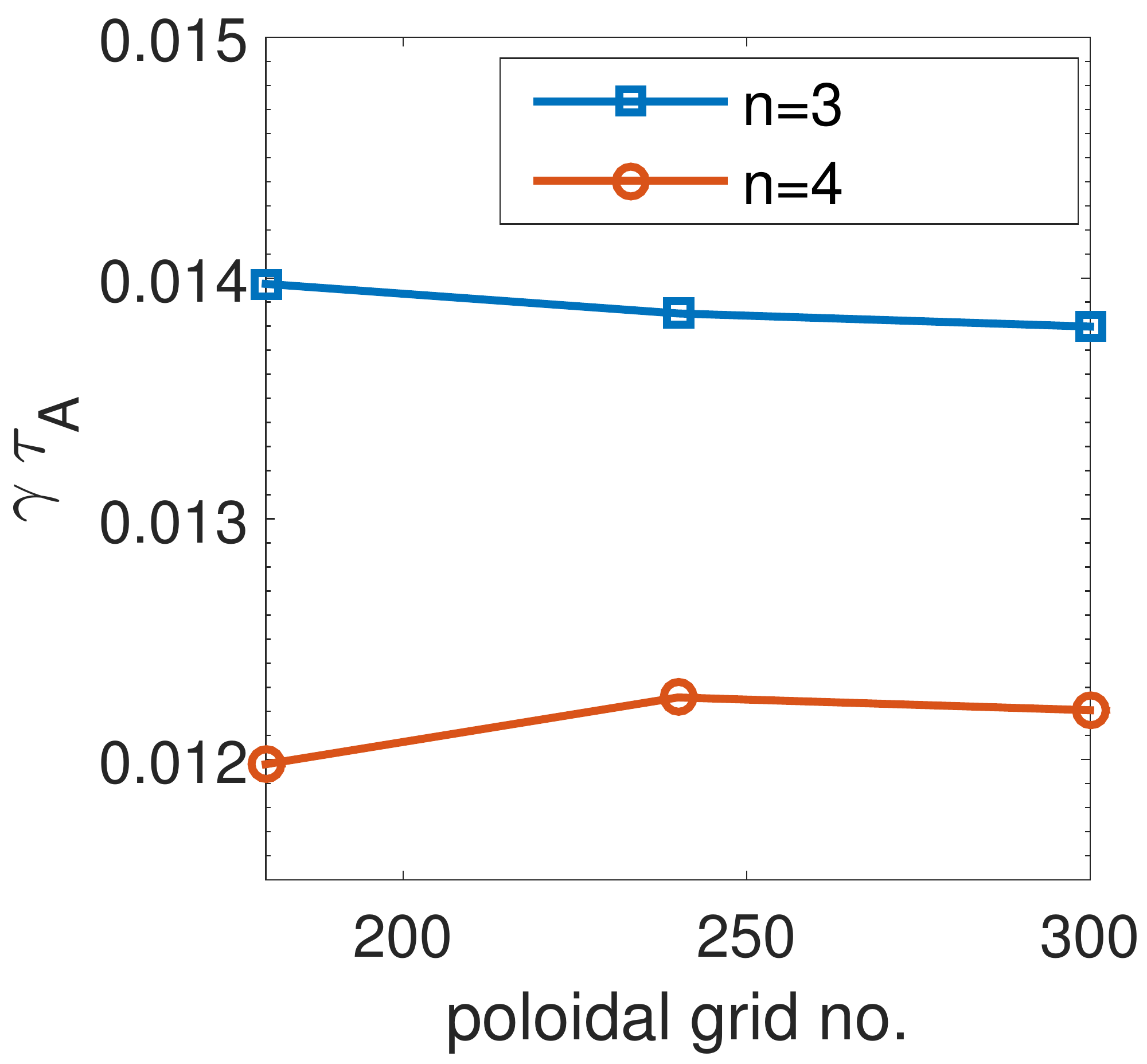} \\
~~~~~~~~~~~~~~~~\\
~~~~~~~~~~~~~~~~\\
\includegraphics[width=7cm]{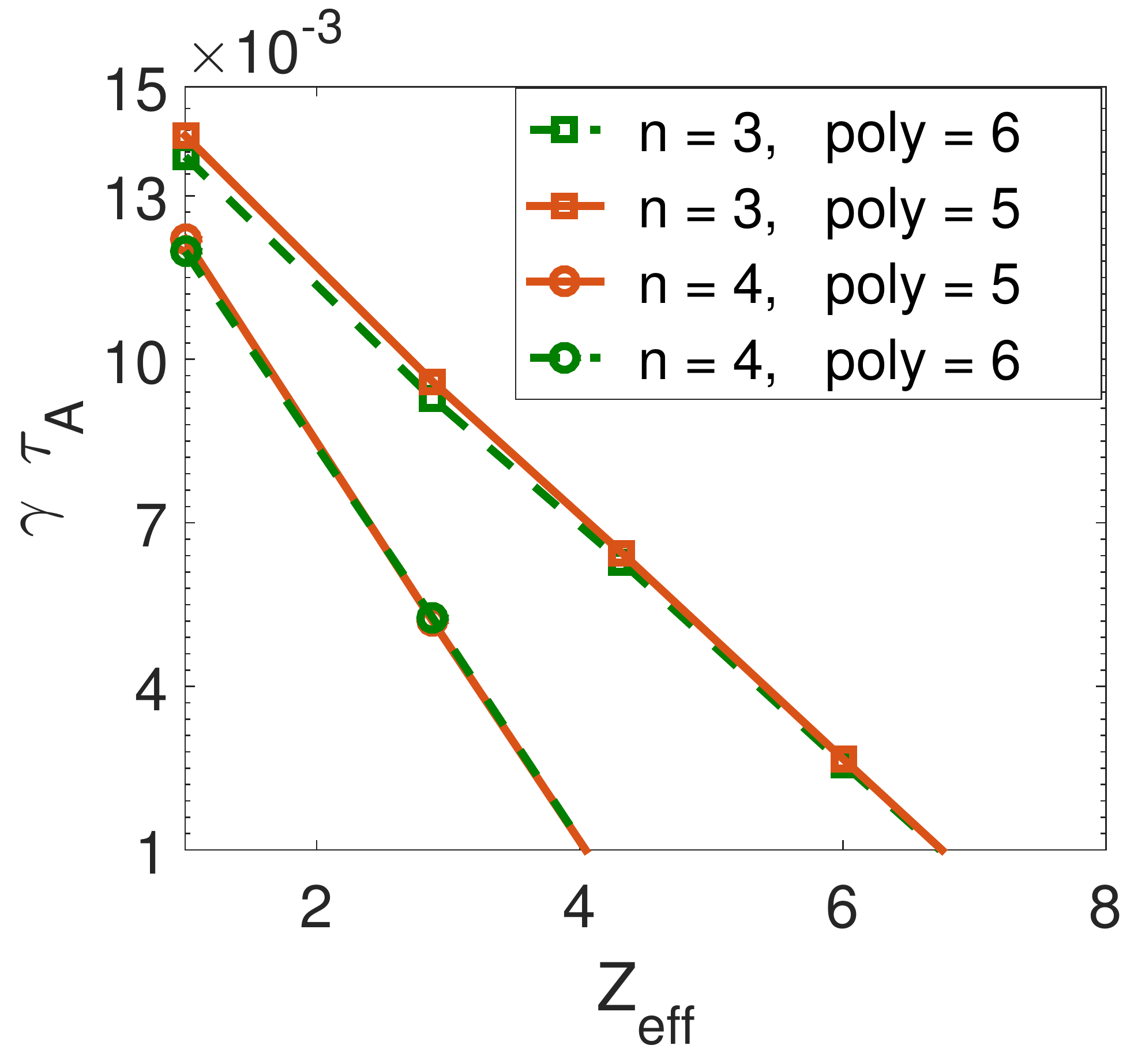}
~~\includegraphics[width=7cm]{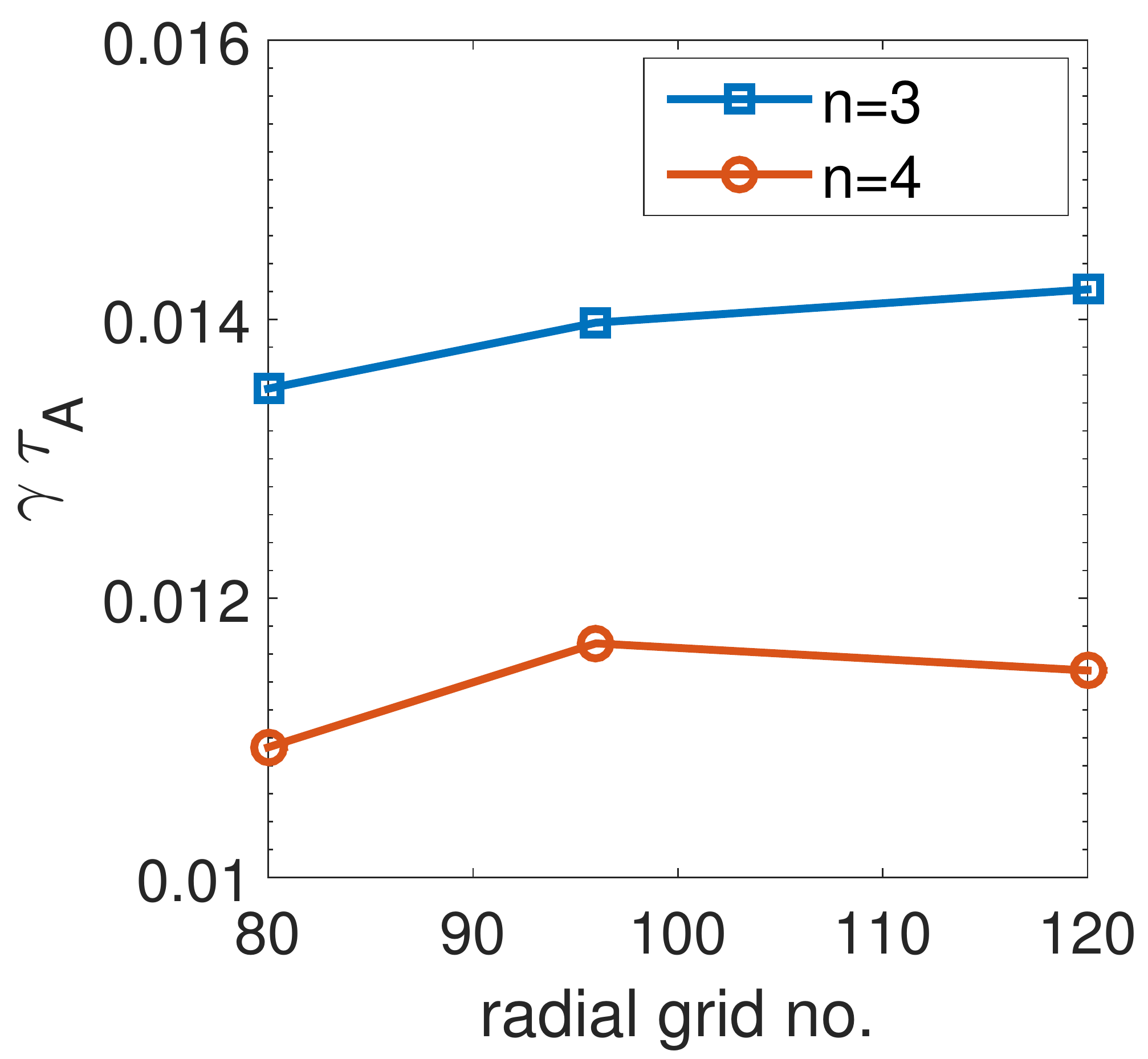}
\caption{\label{conv} (Post-Lithium discharge $\#129038$)~ Linear growth rates for $n=3,4$ modes as function of numerical parameters such as time step, numbers of grid points in azimuthal direction and radial direction, polynomial degree of finite elements used in NIMROD calculations.
}
\end{figure}

In our analysis of the post-lithium case ($\#129038$) using NIMROD,
all toroidal modes in the range $n=1-10$ are found to be stable in the
ideal two-fluid MHD limit, except the modes with $n=3,4$. Modes above
$n=4$ are stabilized after including the two-fluid diamagnetic drift
and the finite Larmor radius (FLR) stabilization effects. Modes
$n=1,2$ are stabilized even in the MHD scenario itself. The left panel
of Fig.~\ref{grpost} shows the two-fluid effects in comparison to the
ideal MHD model. These results are consistent with NSTX ELMy discharge
experiments where the low-$n$ modes are the dominant components in the
ELM precursor signals measured from divertor. Spitzer resistivity
model ($\eta(T_e)=\eta_0 Z_{\rm eff} (T_{e0}/T_{e})^{3/2}$) is adopted
in the calculation, where $T_{e0}$, $\eta_0 Z_{\rm eff}$, $T_e$ denote
electron temperature, resistivity at the magnetic axis, and the
electron temperature profile, respectively. The effect of
gyro-viscosity is also included in the 2-fluid calculations. The whole
simulation domain is divided into a plasma region and a vacuum region
with the real wall configuration of NSTX. The vacuum in these
calculations is modeled using the low temperature, low density and
high resistivity plasma in what is often referred to as halo
region. The values of these parameters in vacuum are naturally
determined by the experimental profiles for both pre- and post-lithium
discharges used in our calculations. A scanning of the resistivity
value through changing $Z_{\rm eff}$ reveals that the growth rates of
$n=3$ and $n=4$ modes decrease with increasing $\eta$, as shown in the
right panel of Fig.~\ref{grpost}. Such a scanning is motivated by the
experiment where $Z_{\rm eff}$ is enhanced substantially
by carbon accumulation as ELMs disappear~\cite{scotti}. The contour
plots of the $n=3, 4$ modes shown in Figs.~\ref{contourpost}
and~\ref{contourpost_zoom} bear the signature of the characteristics
of edge localized modes. The stabilization of the $n=3,4$ modes with
enhanced resistivity in the two-fluid MHD model has not been reported
elsewhere. A thorough test verifying the convergence of above results has been done with respect to time step, azimuthal grid points, and other numerical parameters (Fig.~\ref{conv}).

\begin{figure}[htbp]
\includegraphics[width=7.8cm]{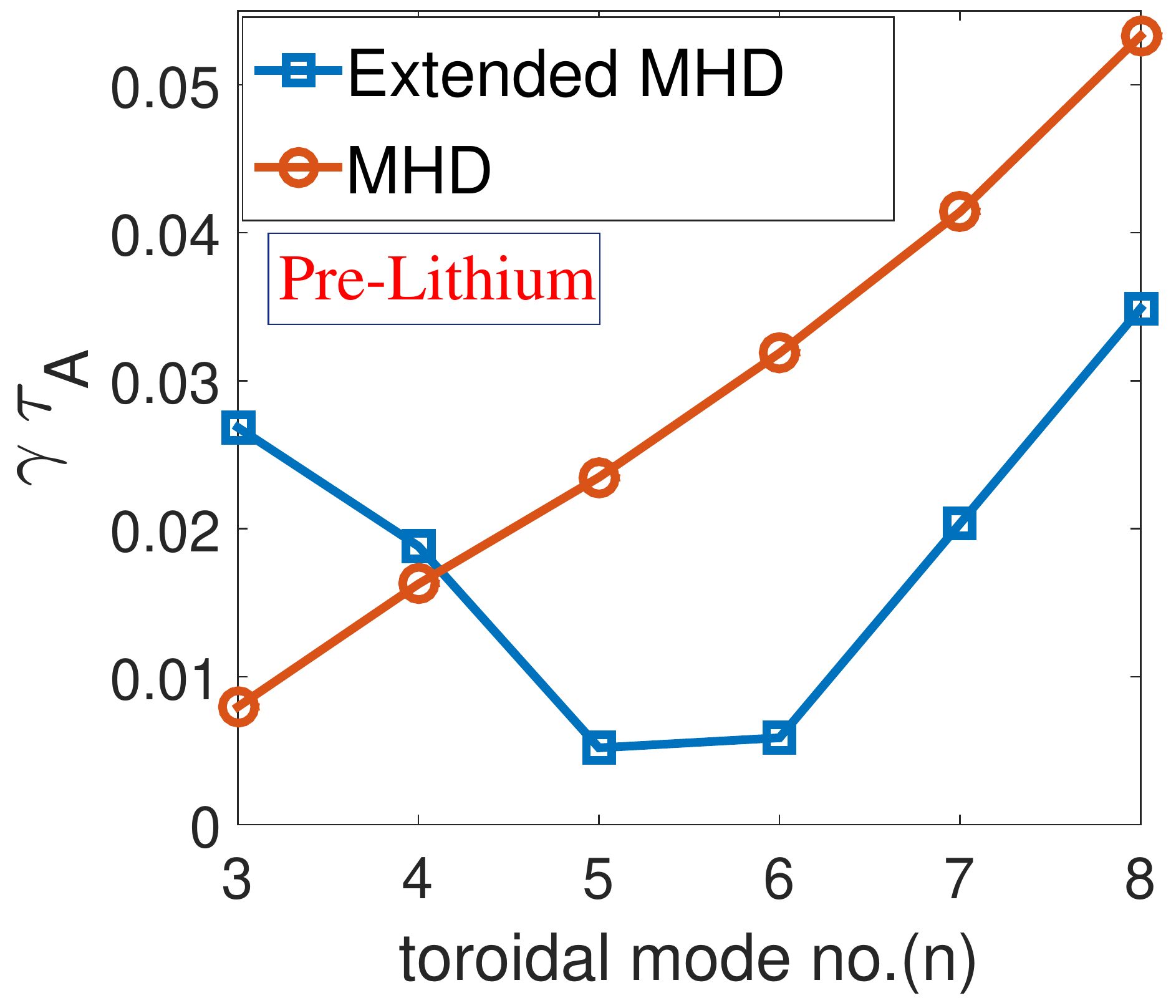}
~~\includegraphics[width=8.0cm]{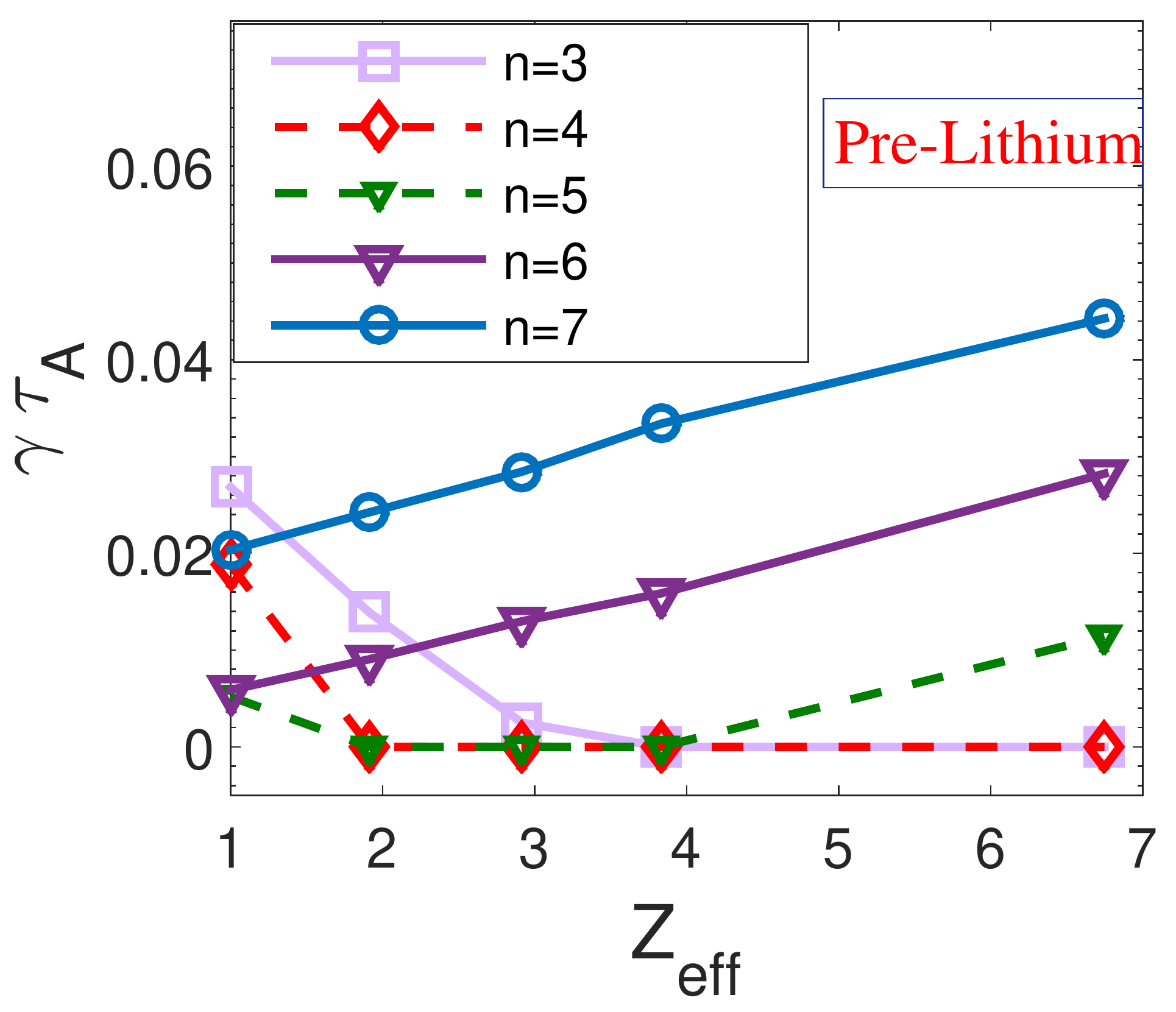}%
\caption{\label{grpre} (Pre-Lithium discharge $\#129015$) ~Left: Linear growth rates as function of toroidal mode number from resistive MHD (blue circled curve) and extended MHD (red squared curve) calculations respectively using NIMROD for the pre-lithium case. Right: Linear growth rates of $n=3,4$ modes as function of $Z_{\rm eff}$.
Experimentally measured $Z_{\rm eff}$ at the edge ranges between 2.0 (at $\psi_N=0.7$) and 2.2 (at $\psi_N=0.9$) in the pre-lithium discharge.}
\end{figure}

It is worth mentioning that this resistive effect on the low-$n$ edge
localized modes is feasible only with the inclusion of the two-fluid
effects in the MHD model. For purely resistive MHD model without
inclusion of 2-fluid effects, the increase in resistivity always
destabilizes all range of the edge localized modes. This contrasts to
the recent study using NIMROD where the resistive stabilizing effects
on low-$n$ edge localized modes appear in both the single-fluid and
the two-fluid MHD models~\cite{jake} which might be arising from
different resistivity model used in ideal and resistive limit. Our
results also differs from resisitive effect studies on
peeling-ballooning modes in other codes as M3D-C$^{1}$ (in single
fluid limit~\cite{ferraro10}) and BOUT$++$ (in 2-fluid limit with
constant resistivity profile \cite{dudson}).  Thus the results
regarding the effect of edge resistivity on low-$n$ peeling-ballooning
modes in a 2-fluid MHD model have not been reported in the
literature, to our knowledge. To check if such an effect also exists
for pre-lithium ELMy H-mode discharge, we have analyzed one of these
cases thoroughly using NIMROD. First, comparison between full
resistive MHD and extended MHD model shows increase of growth rate for
$n=3,4$ modes and reduction of growth rate for other higher $n$ modes
(Left panel of Fig.~\ref{grpre}). This result is consistent with
experimental observation and previous eigenvalue code prediction of
growing low-$n$ modes. A scanning of $Z_{\rm eff}$ through the
resistivity model has been done artificially to evaluate possible
resistive effects on the edge localized modes for the pre-lithium
equilibrium configuration. The results in the Fig.~\ref{grpre} shows
that increase of resistivity has opposite effects for low-$n$
($n=3-6$) and high-$n$ (after $n=6$) modes. That is, the low-$n$ modes
are stabilized with the increase of $Z_{\rm eff}$ up to $4$. However,
high-$n$ modes become more unstable with the increase of resistivity.
But, for the post-lithium case (Fig.~\ref{grpost}),
  modes higher than $n=4$ remain stable even after the increase of
  resistivity.

In summary, the resistive effect towards stabilization of the low-$n$
edge localized modes only from a 2-fluid MHD model is reported, and
such an effect has been applied as a new explanation to the observed
ELM suppression after lithium conditioning in NSTX for the first
time. The conclusion is based on our stability analysis of the
lithium-conditioned ELM-free H-mode experimental profiles from NSTX
using the initial value MHD code NIMROD. For the post-lithium pedestal
profiles, the profile improvement has been found responsible for
suppressing higher-$n$ ($n>4$) modes. However, the $n=3,4$ modes are
found to remain unstable with edge localized mode structure even in
the ideal 2-fluid MHD model. The finding differs from the earlier
eigenvalue-code based analysis and explanation on the probable
mechanism behind the observed suppression of ELMs in the NSTX
experiment~\cite{maingiprl,boyleppcf}.  When the enhanced resistivity
due to the substantial increase of $Z_{\rm eff}$ observed in experiment is
taken into account, the remaining unstable $n=4$ mode can be fully
stabilized along with partial stabilization of $n=3$. This result shows a clear trend of low-n mode stabilization with higher $Z_{\rm eff}$.
Unlike the destabilizing effect of resistivity on all
range of peeling-ballooning modes in the single-fluid MHD model, the
stabilizing effect of resistivity on the low-$n$ edge localized mode
is only obtainable from the two-fluid MHD model. Such a newly found resistive stabilizing effect may contribute to the complete physical mechanism behind the lithium-induced ELM suppression through a virtuous cycle -- the initial refinement in electron pressure profile after the reduction in divertor recycling leads to the initial reduction of ELM frequency and size, with the consequence of increasing $Z_{\rm eff}$ due to impurity accumulation at edge pedestal. The enhancement in $Z_{\rm eff}$ brings in the additional resistive stabilization of the low-$n$ ELMs, which would in turn reinforce the impurity accumulation and the resulting $Z_{\rm eff}$ enhancement until the complete ELM suppression.

To further establish the universality of the effects reported here, more analyses need to be performed for other lithium-conditioned ELM experiments such as those in EAST as well, where the enhancement in $Z_{\rm eff}$ and collisionality are observed in the post-lithium discharge~\cite{huprl}. These may render more detailed understanding of this newly discovered physics, and help further design and implement lithium delivery methods for controlling ELMs in future fusion devices.


The research was supported by the National Magnetic Confinement Fusion Program of China under Grant Nos. 2014GB124002, 2015GB101004, and the 100 Talent Program of the Chinese Academy of Sciences (CAS). Author D.~B. is partially supported by CAS President International Fellowship Initiative (PIFI). We are grateful for the discussions with Charlson C. Kim, the help from Shikui Cheng, and the support from the NIMROD team. This research used resources of the National Energy Research Scientific Computing Center, a DOE Office of Science User Facility supported by the Office of Science of the U.S. Department of Energy under Contract No. DE-AC02-05CH11231. The experiment in NSTX was supported by DOE under Contract No. DE-AC02-09CH11466.
The digital data for this paper is available at http://arks.princeton.edu/ark:/88435/dsp01vx021h49j. 
\bibliography{debu_arxiv}

\newpage

\newpage
%
\newpage

\newpage
 
%
\newpage

\newpage

\end{document}